# Gas-induced segregation in Pt-Rh alloy nanoparticles observed by *in-situ* Bragg coherent diffraction imaging

Tomoya Kawaguchi,[1,†] Thomas F. Keller,[2,3] Henning Runge,[2,3] Luca Gelisio,[2] Christoph Seitz,[2] Young Y. Kim,[2] Evan R. Maxey,[4] Wonsuk Cha,[4] Andrew Ulvestad,[1,§] Stephan O. Hruszkewycz,[1] Ross Harder,[4] Ivan A. Vartanyants,[2,5] Andreas Stierle[2,3] and Hoydoo You[1,*]

[1]*Materials Science Division, Argonne National Laboratory, Argonne, Illinois 60439, USA*

[2]*Deutsches Elektronen-Synchrotron DESY, D-22603 Hamburg, Germany*

[3]*Physics Department, Universität Hamburg, D-20355 Hamburg, Germany*

[4]*Advanced Photon Source, Argonne National Laboratory, Argonne, Illinois 60439, USA*

[5]*National Research Nuclear University MEPhI, 115409 Moscow, Russia*

[†]*Current Address: Institute for Materials Research, Tohoku University*

[§] *Current Address: Tesla, Inc.*

[*] *hyou@anl.gov*

Bimetallic catalysts can undergo segregation or redistribution of the metals driven by oxidizing and reducing environments. Bragg coherent diffraction imaging (BCDI) was used to relate displacement fields to compositional distributions in crystalline Pt-Rh alloy nanoparticles. 3D images of internal composition showed that the radial distribution of compositions reverses partially between the surface shell and the core when gas flow changes between $O_2$ and $H_2$. Our observation suggests that the elemental segregation of nanoparticle catalysts should be highly active during heterogeneous catalysis and can be a controlling factor in synthesis of electrocatalysts. In addition, our study exemplifies applications of BCDI for *in situ* 3D imaging of internal equilibrium compositions in other bimetallic alloy nanoparticles.

Transition metals and their metal alloys have been extensively studied as materials for heterogeneous catalysis and electrocatalysis. In particular, metals in the platinum group and their alloys are among the most important catalysts for organic and electrochemical reactions[1,2]. In industrial applications, they are typically used as nanoparticles to increase efficiency and selectivity. For this reason, the size-shape-activity relationships have been a focus of extensive studies and the importance of surface reactions in nanoparticle surfaces to the relationships has been recognized[3,4]. In some studies, it was further suggested that the shapes and sizes of the nanoparticles can evolve and change during reactions. For example, the surface compositions of bimetallic nanoparticles was reported to change depending on oxidizing or reducing environments[5,6]. Since the changes occur during reactions, studies must be carried out *in-situ* in detail on an individual nanoparticle to fully understand the internal elemental distributions.

A Pt-Rh alloy nanoparticle is an excellent system to study the effect of surface reactions on the compositional redistributions. The unit cell structures of both Pt and Rh are face-centered cubic (fcc) with 3% differences in lattice constants ($a_{Pt}$=3.9242 Å, $a_{Rh}$=3.8034 Å). In bulk, they form solid solutions with no phase separation[7] and negligible compositional ordering[8]. Since there are no significant internal driving forces for compositional redistribution, external driving forces by surface reactions can have a significant effect on Pt-Rh nanoparticles. If the effect of the surface reactions can be detected in a particle with a diameter of ~100 nm, which is roughly the size needed for a suitable signal-to-noise ratio, we expect that the effect must be significant and relevant for real catalyst nanoparticles with higher surface-to-volume ratios.

Bragg coherent diffraction imaging (BCDI) is increasingly popular for imaging of nanoparticles under *in-situ* conditions. [9,10] Real-space 3D images are composed of complex numbers in voxels and the amplitude of a complex number is proportional to the electron density of the crystalline order[11,12], while the phase is proportional to the displacement of the Bragg planes [13-15]. The electron density in principle could encode the composition ratio of Rh and Pt. However, the degree of crystalline order within the particle is generally unknown, and the amplitude is, therefore, unreliable measure of electron density or composition. On the other hand, the *phase* of a BCDI image is a reliable measure of lattice spacings with a high precision and can be used as a means of estimating elemental distributions within a Pt-Rh particle.

The phase of an image reconstructed in BCDI is proportional to the longitudinal component of atomic displacement field, $u_{111}(r)$[15,16]. The lattice spacing deviation from the average, the strain, is obtained from the derivative of the phase image, $\partial_{111} u_{111}(r)/\partial r_{111}$. In the case of electrochemical dealloying at room temperature[16], the deviation was used to map the mechanical strain induced by the dealloying process. In the case of an alloy nanoparticle at high temperatures, however, the alloy elements can diffuse freely and the mechanical strain can be replaced by the *compositional strain*, i.e., the strain (lattice spacing change) induced by the compositional heterogeneity.[17] Then, the local composition can be determined from the compositional strain via Vegard's law assuming that the changes in local composition isotopically expand/contract the cubic unit cell. Thus, we will present our results as 3D images of local Rh composition converted from the phase [17] and 3D images of normalized electron density.

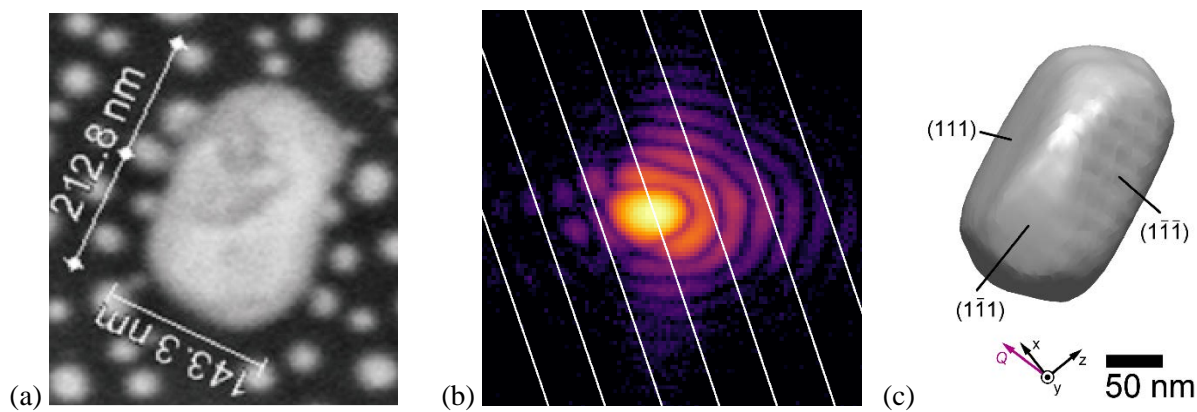

Fig. 1. (a) An SEM image of the $Pt_{2/3}$-$Rh_{1/3}$ alloy nanoparticle. The smaller nanoparticles in the SEM image are pure Rh particles. (b) A 2D coherent diffraction pattern of the 111 peak from the particle in a logarithmic intensity scale. The white equi-$2\theta$ lines indicate the expected Bragg peak maximum for every 10% change in the mean Rh composition of the particle. (c) A 3D reconstruction of the same particle using BCDI in a top-down view as is in (a). *x, y,* and *z* of the laboratory coordinate system correspond to the outboard direction of the storage ring, the vertical direction, and the incident x-ray direction, respectively.

Fig. 1 shows a scanning electron microscopy (SEM) image of the sample nanoparticle with a nominal $Pt_{2/3}$-$Rh_{1/3}$ composition (a), a slice of Bragg diffraction data from the particle (b), and a 3D shape reconstructed from the data (c). The sample was prepared by depositing a Rh layer at 630 °C over Pt nanoparticles dewetted on $Al_2O_3$(0001) substrate.[17] Among several particles examined and tested, the nanoparticle shown in Fig. 1(a) was selected for detailed studies. A hierarchical fiducial makers were used to identify the same particle with a SEM in the laboratory[18] and with a confocal microscope in the beamline. The 3D shape (c) and 3D phase were reconstructed by iterative phase retrieval algorithms standard for BCDI[18,19] in combination with a guided genetic algorithm[20] after applying flat-field correction on the detector and subtracting background scattering. The voxel size of BCDI images was 6.7 nm and the resolution was ~13 nm. The x-ray induced oxidation was negligible, consistent with prior observations[21].

The BCDI measurements were carried out under variable gas environments at atmospheric pressures, pure He, He with 2.7%$O_2$, He with 5%$O_2$, and He with 3.8%$H_2$, at two temperatures, 550 and 700 °C. The temperature was calibrated with substrate peak $2\theta$ position. The environments were chosen for neutral, oxidizing, and reducing conditions and the two temperatures were chosen for Rh diffusion lengths. Rhodium atoms in Pt matrix are expected to diffuse 6 nm and 70 nm at these temperatures, respectively,[22,23] over the diffusion time of $10^3$ sec, which is close to the approximate wait time at each gas change. The diffusion length at 550 °C was not sufficient to achieve a full thermal equilibrium over the entire particle but was sufficient for composition redistributions to be seen in BCDI measurements. At the elevated temperature of 700 °C, the particle-average lattice spacing decreased significantly over time,[17] indicating the alloy nanoparticles slowly incorporated Rh atoms from neighboring Rh nanoislands during the long experiment because of the faster Rh diffusion. The Rh incorporation was not significant at 550 °C but the lattice spacing responded whenever the gas flows were changed. The measurements at 400 °C or below showed little effect of gas environments[17] as expected from the extremely short thermal diffusion lengths of 0.15 nm for $10^3$ sec diffusion and are not discussed here.

The accurate Bragg scattering angle was compared to the angles expected from the thermal expansion rates of the Pt-Rh alloys known in the literatures[24,25]. This provided us a means of determining the particle average composition precisely for any given environment and temperature. The initial average composition at 550 °C determined in this way was close to our target composition of $Pt_{2/3}Rh_{1/3}$ although the composition evolved continuously depending on the environment and temperature during the course of the experiments. [17]

Fig. 2 shows the 3D images of normalized amplitude (a) and composition difference (b) reconstructed at 550 °C. The $H_2$ exposure significantly decreased the electron density of the particle as indicated by the irregularities in the isosurfaces shown in Fig. 2(a), while the exposures of He and $O_2$ have almost no effect on the particle shape. While the density irregularity suggested an increased Rh composition, it can also indicate a significant reduction in crystalline order [11,12] as discussed earlier, and the composition image was obtained from the phase image. Fig. 2(b) shows the cross-sectional views of the difference images in Rh composition obtained from the phase.[17] In an oxidizing atmosphere ($O_2$), only subtle growth of the Pt-rich region (green region) was observed near the surface. A reducing atmosphere ($H_2$), however, significantly affected the composition, as Rh-rich regions (purple region) appeared near the surface and composition variations within the particle became significant.

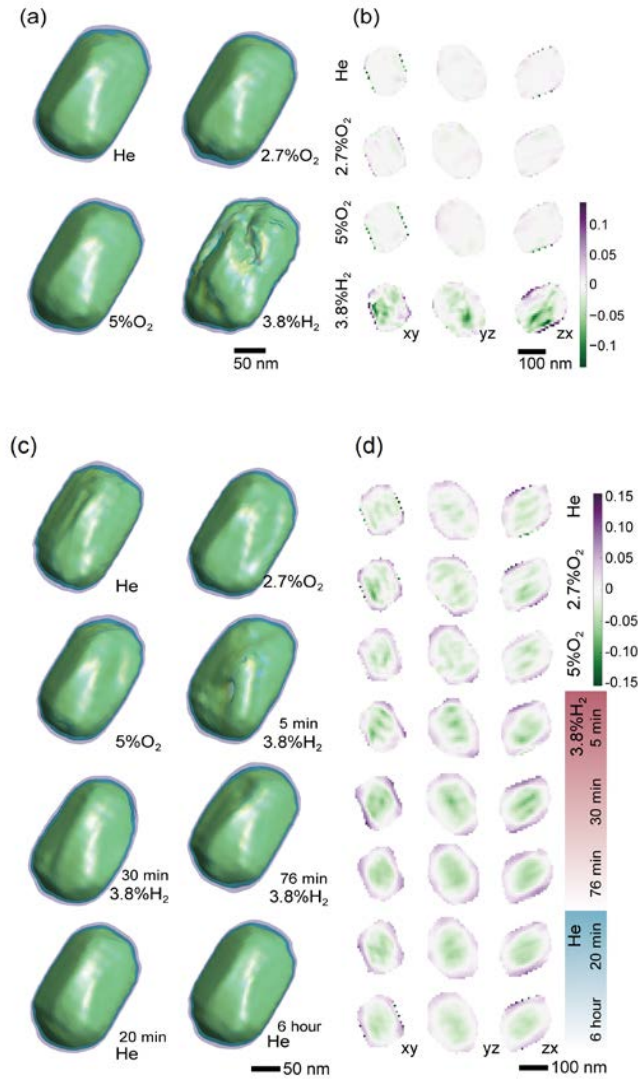

Fig. 2. (a,c) Electron density normalized by the maximum amplitude and (b,d) Rh composition difference from the particle average, obtained from compositional strain at various gas atmospheres at 550 °C (a,b) and at 700 °C (c,d). (a,c) Translucent isosurfaces at 45% (blue envelope), 65% (green shell) and 85% (yellow core) of the mean electron density are shown. (b,d) The color bar indicates the Rh composition difference ($\Delta x_{Rh} = x_{Rh} - \bar{x}_{Rh}$)[17] where $\bar{x}_{Rh} \simeq 0.33$ and $x_{Rh}$ ranges from ~0.13 (green) to ~0.48 (purple) depending on the exact value of the particle average, $\bar{x}_{Rh}$.

The particle shape obtained from the BCDI remained largely unchanged at 700 °C and electron density isosurfaces in Fig. 2(c) appeared not sensitive to gas environments. However, upon close inspection of the images in Fig. 2(c), some shape change can be seen in the $H_2$ atmosphere at the initial 5 min exposure of 3.8%$H_2$. The shape was restored in the images for the 30 and 76 min exposures, presumably as the system reached a steady state or a full thermal equilibrium because of the faster diffusion of Rh at this temperature. The compositions, shown in Fig. 2(d), indicate a significant increase of Rh compositions near the surface in all gas environments.

The images were further analyzed to obtain the radial distributions of the composition, plotted in Fig. 3 as a function of the distance from the 45% isosurface defined in Fig. 2. The radial distance of a voxel was defined as the distance *from* the nearest voxel on the 45% isosurface. Then, the radial distance up to 60 nm measured from the surface was divided to 25 shells and the Rh compositions were averaged at each shell. The zeros of the Rh composition were set to the compositions averaged for the core of the particle, defined by the radial distance from the surface larger than 40, to show clearly the differences at the surface.

In Fig. 3(a), the radial distribution analysis for the 500 °C images clearly demonstrates the sensitivity of the surface composition to the gas environments. In 5%$O_2$ environment, the Pt-rich (Rh-poor) region appeared at the bins near the surface ($<$ 15 nm) while the composition is relatively flat in He and in 2.7%$O_2$. At the same time, the average composition (inset) decreased slightly under 2.7%$O_2$ atmosphere, which continued under 5% $O_2$. The decrease of the average Rh composition continues (inset) and Rh composition depletes further near the surface (◊). On the other hand, the H$_2$ atmosphere yields the 6% *increase* of Rh-composition at the surface as well as the considerable increase of the average Rh composition.

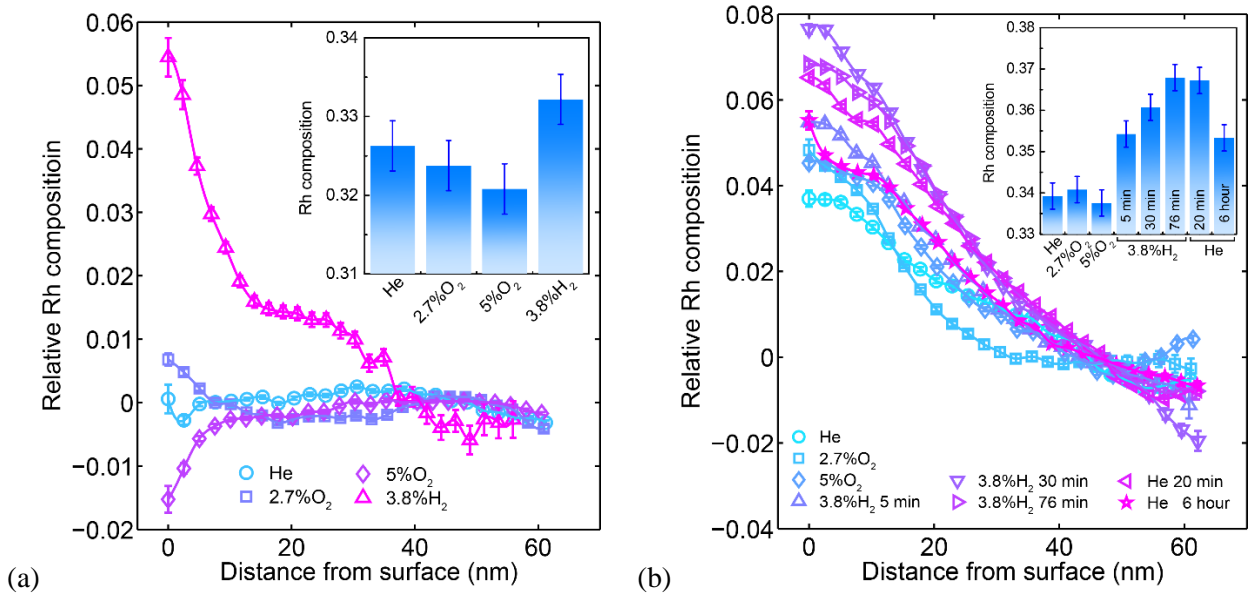

Fig. 3. (a) Radial distributions of the relative Rh composition converted from compositional strain at 550 °C. (b) Radial distributions of the relative Rh composition converted from compositional strain at 700 °C. The zero of the relative Rh composition was set to the mean value of the composition in the core of 40 nm or larger from the surface. The displayed time in (b) is the elapsed time after the gas introduction. The insets exhibit the particle average compositions of Rh determined from the Bragg angles. These values are used to set the zeros in the vertical scale bars in Fig. 2. Error bars in the inset was estimated based on the pixel size of the 2D detector.

In Fig. 3(b), the composition slope with Rh increased near the surface is evident at 700 °C. Even under O$_2$ flow, Rh was enriched near the surface. The trend of the Rh surface enrichment was enhanced in H$_2$ (Δ, ∇, ▷), consistent with the results at 550 °C. The increased Rh composition at the surface also led to a steady increase of the overall Rh composition as shown in the inset. When the gas was switched back to He (◁ and ☆), however, the Rh-rich surface region and the overall Rh composition decreased back slowly with time. In He flow, the composition slope decreased but still remained even after 6 hours.

Our results show that compositional distribution of Pt-Rh alloy nanoparticles is dynamic at high temperatures and sensitive to the environment condition. Some rationale to the observed behavior is illustrated in Fig. 4. Since the sample was exposed to air after preparation, the particle surface was initially covered by native oxides of rhodium because rhodium can form crystalline oxides such as hexagonal $Rh_2O_3$ or rutile $RhO_2$ while platinum does not easily form crystalline oxides.[26] The native oxides are expected to be extremely thin, limited to a few atomic layers, as illustrated in Fig. 4(a) since the sample is cooled to room temperature before exposure to air. Note that any diffraction from the oxides, thin or thick, unlikely contribute to the 111 Bragg peak measured here because of its different crystalline structure. Nonetheless, Rh oxide formation is expected and can be a driving force for the observed compositional redistribution in oxidizing environments. [6,27]

When the particle is heated in the $O_2$ atmosphere, Rh, being more oxophilic than Pt,[28] forms oxides faster and thicker than Pt does. This causes Rh dealloyed at the surface of the particle by forming Rh oxides, which eventually yields the Rh-poor, Pt-rich region on the surface as in Fig. 4(b). In the $H_2$ atmosphere, the reverse reaction is expected. The Rh oxide is reduced and Rh atoms are incorporated to the particle. This simple oxidation-reduction model does not completely explain the results that the Rh was significantly enriched in the $H_2$ atmosphere. If the Rh oxides formed in $O_2$ simply incorporated back to the particle in $H_2$, the Rh composition on the surface could not have increased as significantly as observed.

The composition increase of Rh at the surface was 0.06, as shown in Fig. 3(a). The actual surface composition profile can be much narrower and higher because the distribution is a convolution of the actual distribution with the BCDI resolution (13 nm). Such significant surface concentration can occur only if external Rh atoms, diffused from neighboring Rh nanoislands, are incorporated to the surface of the particle. Although we have no direct evidence, it is likely even at 550 ºC that Rh atoms shuttle back and forth from the neighboring Rh nanoislands surrounding the Pt-Rh particle of this study. The Rh migration explains naturally the Rh-rich surface. This also increases the overall Rh composition as illustrated in Fig. 4 (c), which could not have taken place without the Rh migration.

At the relatively high temperature of 700 °C, the situation may be different. The equilibrium oxygen partial pressure of the Rh oxidation is quite high at ~$10^{-2}$ atm[28], which means that Rh oxides tend to spontaneously reduce to metal in the He atmosphere. Thus, alloying of Rh by the Rh migration can continue at this temperature without the Rh oxide on the surface in He atmosphere as well as in $H_2$ atmosphere. As discussed earlier, the diffusion length ($\sqrt{Dt}$) of Rh at this temperature is 70 nm at $10^3$ sec that is as large as the average radius of the particle, allowing the entire particle to reach a full thermodynamic equilibrium whenever the gas environment is changed. This explains why the radial distributions appear all similar in the different gases. However, it still shows significant changes of the Rh composition near the surface depending on the gas flow, much like the case at 550 °C. The continuous Rh migration from the Rh nanoislands explains the overall slope of the gradual decrease of Rh from the surface to the core. However, it is also possibly driven thermodynamically towards the truncated surface[3] as experimentally observed in Rh-Pt nanoparticles[29], which requires further theoretical clarification.

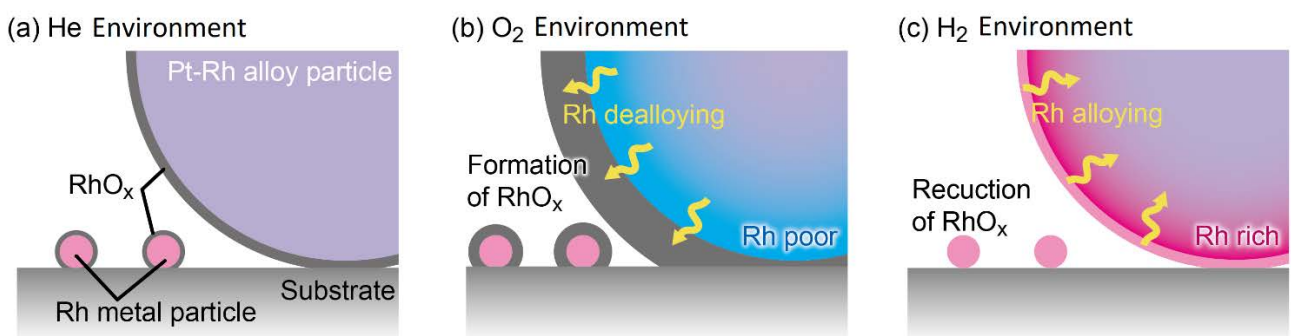

Fig. 4. Schematics of alloying and dealloying processes in He, $O_2$, and $H_2$ environments. (a) In He, the surfaces of the Pt-Rh particle and Rh nanoparticles all have a thin native oxide layer. (b) In $O_2$, Rh oxide layer grow faster because Rh is more oxophilic than Pt. (c) In $H_2$, the Rh oxide layer reduces, which increases the Rh compositions at the surface.

An additional interesting observation was a preferential emergence of Rh-rich {111} facets in He and $O_2$.[17] This indicates that anisotropic, oxidation induced Rh segregation can take place even accompanied by a shape change, more likely for smaller particles. It is of interest to experimental and theoretical catalysis studies to further explore the particle shape and composition evolutions of the alloy particles at various environments and temperatures.

In conclusion, the BCDI technique, highly sensitive to nanoscale variations of lattice constants, is shown to be a powerful technique in studying *in situ* 3D internal composition map and dynamic shell-core compositional rearrangement in a $Pt_{2/3}Rh_{1/3}$ alloy nanoparticle. At the intermediate temperature of 550 ºC, the surface Rh composition significantly decreases in an oxidizing environment and increases in a reducing environment. At higher temperatures, the Rh composition is significantly higher on the surface regions and gradually decreases towards the core of the particle regardless of the gas environments. However, the composition distribution shows the gas dependences similar to the results at 550 ºC. The redistributions of metals observed in alloy nanoparticles as large as ~100 nm offers an insight into the dynamic nature of the generally much smaller nanoparticle catalysts in practical usages, in particular for gas-phase catalysis and in fuel-cell catalysts operating at elevated temperatures, suggesting that the active compositional redistribution and other structural dynamics must be taken into account for rational design of new catalysts and electrocatalysts. This study also open a possibility of composition imaging studies on other binary alloys in equilibrium under various extreme environmental conditions.

### Acknowledgements

The work at Argonne, x-ray measurements and data analysis (TK AU SH HY), was supported by the U.S. Department of Energy (DOE), Office of Basic Energy Science (BES), Materials Sciences and Engineering Division, and x-ray work (WC RH) and use of the APS, by DOE BES Scientific User Facilities Division, under Contract No. DE-AC02-06CH11357. The work at DESY, x-ray measurements and sample preparation and marker-based nano-transfer (TFK HR LG IV AS) was supported by the EU-H2020 research and innovation program under grant agreement No 654360 NFFA-Europe. The use of the FIB dual beam instrument granted by BMBF (5K13WC3, PT-DESY) is acknowledged. We acknowledge marking the regions of interest by IBID/EBID and the SEM post-analysis by Satishkumar Kulkarni and Arno Jeromin (DESY NanoLab). One of the authors (TK) thanks the Japanese Society for the Promotion of Science (JSPS) for JSPS Postdoctoral Fellowships for Research Abroad.